\documentclass[amsmath,twocolumn,superscriptaddress,prl,aps]{revtex4}

\usepackage{epsfig}
\usepackage{graphicx,color}
\usepackage{dcolumn}
\usepackage{bm}
\usepackage{amssymb}
\usepackage{xspace}

\begin{document}


\title{Spontaneous quantum Hall effect via thermally induced quadratic Fermi point}

\author{Gia-Wei Chern}
\affiliation{Theoretical Division, Los Alamos National Laboratory, Los Alamos, New Mexico 87545, USA}
\affiliation{Department of Physics, University of Wisconsin, Madison, Wisconsin 53706, USA}

\author{C. D. Batista}
\affiliation{Theoretical Division, Los Alamos National Laboratory, Los Alamos, New Mexico 87545, USA}

\date{\today}

\begin{abstract}
Gapless electronic systems containing topologically nontrivial Fermi points are sources of various topological
insulators. Whereas most of these special band-crossing points are built in the electronic structure of the
non-interacting lattice models, we show that a quadratic Fermi point characterized by
a non-zero winding number emerges with a collinear triple-${\bf Q}$ spin-density-wave state that 
arises from a perfectly nested but topologically trivial Fermi surface.
We obtain a universal low-energy Hamiltonian for the quadratic Fermi point and show
that such collinear orderings are unstable against the onset of  scalar spin chirality  that opens a gap and induces a
spontaneous quantum Hall insulator as the temperature tends to zero.
\end{abstract}

\maketitle

Fermi points with nontrivial Berry connection have recently been the focus of intense theoretical 
and experimental effort. 
These singular points are the momentum-space counterpart of topological defects in ordered media;
both are characterized by a topological invariant~\cite{volovik03}. 
Topological Fermi points are usually robust against small perturbations which preserve certain discrete symmetries 
of the system. On the other hand, when a gap opens at a singular Fermi point, its topological nature
can be transferred to the resultant insulating state, giving rise to intriguing phases such as spontaneous quantum  Hall 
or quantum spin-Hall states \cite{hasan10}.

The most famous  Fermi point  is the Dirac point. The Fermi ``surface'' of a half-filled
honeycomb lattice, like graphene,  consists  of a pair of these points. Each Dirac point is similar to a vortex in the Brillouin zone (BZ) and is characterized by a topological winding number~$n=\pm 1$~\cite{volovik03}. Haldane showed that the inclusion of complex hoppings in a honeycomb lattice  opens up a gap at both Dirac points, giving rise to a quantum Hall state in absence of external magnetic field ~\cite{haldane88}. Similar ideas for realizing quantum Hall insulators through gapped Dirac points have been 
explored in double-exchange models on square, kagome, and checkerboard lattices~\cite{chen10,ohgushi00,venderbos12}.
By creating and manipulating Dirac points with the aid of artificial gauge fields, it is also possible
to control topological phase transitions in optical lattices~\cite{goldman09,bermudez10}.

Quantum Hall effect can also be induced by opening a gap at a topological Fermi point with a quadratic
band dispersion~\cite{sun09}. These points are vortices carrying multiple topological charges $n = \pm 2$ 
in momentum space~\cite{heikkila10,note1}.  Because of its higher winding numbers, a quadratic Fermi point can decay into
several elementary Dirac points while preserving the total topological charge~\cite{heikkila10}.
Unlike Dirac points where interaction effects are suppressed due to a vanishing density of states (DOS),
quadratic Fermi points are unstable against arbitrarily weak short-range
interactions due to their finite DOS~\cite{sun09,uebelacker11}.
These  Fermi points are usually protected in lattice models by $C_4$ or $C_6$ point group symmetries. Explicit examples are tight-binding models on checkerboard and kagome lattices, respectively~\cite{sun09,liu10}. 
Topological quadratic Fermi points also appear in physical systems such as bilayer 
graphene~\cite{vafek10,manes07,martin08a,dietl08,nandkishore10,min08}, 
photonic crystals~\cite{chong08}, oxide heterostructures~\cite{banerjee09}, and surface state of topological insulators~\cite{fu11}.

\begin{figure}[t]
\includegraphics[width=0.85\columnwidth]{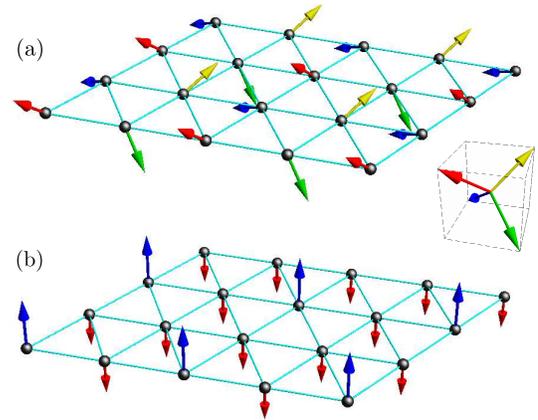}
\caption{(Color online) (a) Noncoplanar spin order with a quadrupled unit cell on a triangular lattice. 
The magnetic moments at the four sublattices point toward corners of a regular tetrahedron (see the inset).
(b) Collinear spin order on the triangular lattice. In the enlarged unit cell,
one site has large spin moment $3\bm\Delta$ while the other three sites have small moment $-\bm\Delta$.
\label{fig:sdw}}
\end{figure}

In this paper we present the first example of a quadratic Fermi point with non-zero winding numbers that emerges from 
a topologically trivial Fermi surface through electron interactions. 
This special Fermi point results from a collinear triple-${\bf Q}$
magnetic ordering that is favoured by perfect nesting of the original Fermi surface.  
Moreover, by analyzing the stability of this emergent Fermi point, we demonstrate that a non-coplanar chiral spin order 
is always more stable than any other triple-${\bf Q}$ ordering (collinear or coplanar) as temperature tends to zero. 
The simple reason is that the non-coplanar state lowers its energy by  opening a gap  at the emergent Fermi point that 
also leads to {\it spontaneous quantum Hall effect}. To illustrate these phenomena we consider Hubbard-like models 
on 2D lattices with a $C_6$ symmetry such as triangular or honeycomb lattices. 
It has been shown recently  that the spin-density-wave (SDW) ground state of such 
models has a noncoplanar chiral order when the chemical potential reaches the saddle-point of the band 
structures~\cite{martin08,wang12,li10,li11,yu11}. This magnetic ordering has a quadrupled unit cell with local spins
pointing toward corners of a regular tetrahedron; an example of such ordering on the triangular lattice
is shown in Fig.~\ref{fig:sdw}(a).

The spontaneous quantum Hall effect of the SDW state arises from a non-zero scalar
spin chirality $\langle\mathbf S_i\cdot\mathbf S_j \times \mathbf S_k \rangle$ of the underlying magnetic 
structure \cite{martin08,shindou01}.
When conduction electrons propagate through such a spin texture, their wavefunction acquires a Berry phase which is equal to
half the solid angle subtended by local moments of each elementary plaquette.
The Berry flux which is indistinguishable from a real magnetic flux induces the quantum Hall effect.
Indeed, quantized Hall conductivity $\sigma_{xy} =  e^2/h$ has been explicitly calculated for
the chiral SDW state in both triangular and honeycomb lattices~\cite{martin08,li10,li11,shindou01}.

Here we show that the recently discovered finite-temperature SDW phase with {\it collinear} magnetic 
moments~\cite{nandkishore12} provides a useful starting point for understanding the origin of  the spontaneous quantum Hall effect 
that necessarily appears at lower temperatures. 
A striking feature of this collinear SDW state shown in Fig.~\ref{fig:sdw}(b) 
is the existence of a quadratic Fermi point in its mean-field band structure~\cite{nandkishore12}. 
Gapless charge excitations in this SDW state exist only for one spin branch. 
More importantly, we show that the quadratic Fermi point has a topological winding number $n = \pm 2$.
We derive a universal low-energy Hamiltonian for excitations around the Fermi point.
Remarkably, the gap of the quantum Hall state is proportional to the scalar spin chirality of the ordered spin state. 
Consequently, the transition from collinear to chiral SDW order coincides with the onset of quantum Hall effect.

\begin{figure}[t]
\includegraphics[width=0.99\columnwidth]{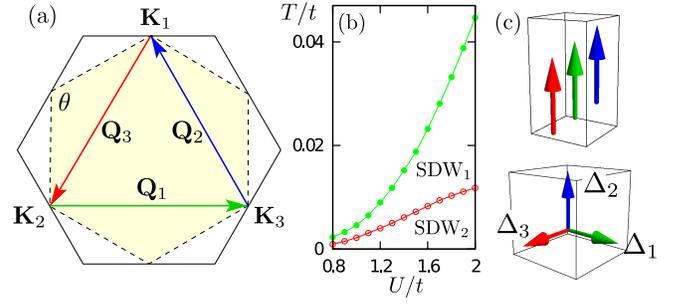}
\caption{(Color online) (a) The Brillouin zone of lattices with a $C_6$ symmetry and the Fermi surface
when electron filling reaches the Van Hove singularity of the band structure. The three nesting wavevectors
are $\mathbf Q_1 = (2\pi/\sqrt{3},0)$ and $\mathbf Q_{2,3} = (-\pi/\sqrt{3},  \pm \pi)$. (b)~Phase diagram of
the triangular-lattice Hubbard model. SDW$_1$ and SDW$_2$ denote spin-density-wave states with collinear
and noncoplanar magnetic moments, respectively. (c)~Configurations of the
three vector order parameters $\bm\Delta_\eta$ in the collinear (top) and chiral `tetrahedral' (bottom) SDW phase.
\label{fig:phase}}
\end{figure}

We start by considering  SDW instabilities in 2D lattices with a $C_6$ symmetry, such as  
triangular, honeycomb, kagome and their decorated variants~\cite{ruegg10}.
The tight-binding DOS  includes one or two Van Hove singularities.
For example, a logarithmically divergent DOS appears for filling factor 3/4 in the triangular lattice, 
and 3/8 or 5/8  in the honeycomb lattice.  At these filling factors, the Fermi surface
is a regular hexagon inscribed within the BZ; see Fig.~\ref{fig:phase}(a). 
Remarkably, pairs of parallel edges of this Fermi surface are perfectly nested by 
wavevectors~$\mathbf Q_\eta$ which are equal to half of reciprocal lattice vectors.
The perfect Fermi surface nesting is quite robust and is broken only by the inclusion of third nearest-neighbor or longer range hopping amplitudes.

SDW order appears as a weak-coupling instability induced by perfect Fermi surface nesting.
In particular, the nature of the SDW instability is mainly controlled by the states that are close to the saddle points 
${\mathbf K}_{\eta}$ shown in Fig.~\ref{fig:phase}(a) ($\eta = 1,2,3$), where a vanishing Fermi velocity gives rise to a logarithmically divergent DOS.
The effective Hamiltonian for the SDW ordering expressed in terms of these low-energy electrons is~\cite{nandkishore12}
\begin{eqnarray}
	\label{eq:H_sdw}
	H &=& \sum_{\eta; \alpha; {\mathbf k}} \varepsilon_{\eta}(\mathbf k)\, c^{\dagger}_{\eta \alpha, \mathbf k}
	c^{\phantom{\dagger}}_{\eta\alpha,\mathbf k} 
 	\\
	&-& \sum_{\xi\neq\zeta; \alpha;\beta} \sum_{{\mathbf k} ; {\mathbf q}; {\mathbf p}} 
	g_{\alpha,\beta} \,c^{\dagger}_{\xi\alpha,{\mathbf k-\mathbf q}} c^{\dagger}_{\xi\beta, {\mathbf k+\mathbf q}}
	c^{\phantom{\dagger}}_{\zeta\beta, {\mathbf k+\mathbf p}} c^{\phantom{\dagger}}_{\zeta\alpha,{\mathbf k-\mathbf p}}  
\nonumber \\
	&+&  \sum_{\xi\neq\zeta; \alpha;\beta} \sum_{{\mathbf k} ; {\mathbf q}; {\mathbf p}}  
	g'_{\alpha,\beta} c^{\dagger}_{\xi\alpha, {\mathbf k-\mathbf q}} c^{\dagger}_{\zeta\beta, {\mathbf k+\mathbf q}}
	c^{\phantom{\dagger}}_{\zeta\beta,{\mathbf k+\mathbf p}} c^{\phantom{\dagger}}_{\xi\alpha,{\mathbf k-\mathbf p}}~, \nonumber
\end{eqnarray}
where $c^\dagger_{\eta \alpha,\mathbf k}$ creates an electron with spin $\alpha \!=\,\uparrow,\downarrow$
and momentum ${\mathbf K}_{\eta}+{\mathbf k}$ ($|{\mathbf k}| \ll 1$), while $g$ and $g'$ denote the forward and umklapp scatterings,
respectively~\cite{nandkishore12}.  The dispersion around the three saddle points is given by $\varepsilon_{\eta}(\mathbf k)
= k_{\xi} k_{\zeta}/m^*$, where $1/m^* \sim \mathcal{O}(t)$ is of the order of bandwidth, $(\eta\xi\zeta)$ 
is a cyclic permutation of (123), and $k_\xi = \mathbf k \cdot {\hat {\mathbf e}}_\xi$, with $\hat{\mathbf e}_3 = (1,0)$  
and $\hat{\mathbf e}_{1,2} = (-\frac{1}{2},\,\pm\frac{\sqrt{3}}{2})$. For the special case of Hubbard model, we have $g = g' = U \delta_{\alpha, {\bar \beta}}/N$, where $U$ is the on-site Coulomb repulsion, $\bar \beta$ denotes the opposite spin, and $N$ is the total number of sites.
Although renormalization-group (RG) analysis indicated that superconductivity instability is asymptotically dominant \cite{Nandkishore11},
the SDW vertex is the largest at intermediate RG scale and becomes dominant by slightly
doping away from the Van Hove singularity \cite{Kiesel11}.

The effective filling fraction restricted to this low-energy model can be obtained by a simple geometrical consideration. 
As shown in Fig.~\ref{fig:phase}(a), the filled region in the vicinity of a saddle point is bounded by two straight lines 
with an angle $\theta=2\pi/3$ between them.  This gives rise to an effective filling factor: $\nu_{\rm eff} = \theta/\pi = 2/3$.
In the folded BZ of the triple-$\mathbf Q$ SDW states, the three saddle points are shifted to the zone center;
each corresponds to two electron bands with different spin species. An effective 2/3 filling indicates that
4 out of the 6 bands are filled in the insulating state.

The component of the SDW order parameter associated with the nesting wavevector $\mathbf Q_\eta$ is
$
	\bm\Delta_{\eta} = \frac{v}{3} \sum_{\mathbf k}\, \bigl\langle c^{\dagger}_{\xi \alpha,\mathbf k}\,
	\bm\sigma^{\phantom{\dagger}}_{\alpha\beta} c^{\phantom{\dagger}}_{\zeta \beta,\mathbf k} \bigr\rangle,
$
where $v = g + g'$. A mean-field decoupling of the interaction terms in (\ref{eq:H_sdw}) based on this SDW order parameter 
yields a Hamiltonian:
\begin{eqnarray}
	\label{eq:H_MF1}
	H_{\rm MF} = \frac{1}{v}\int d\mathbf r \sum_{\eta} \left|\bm\Delta_\eta\right|^2 
	+ \sum_{\xi\zeta}\sum_{\alpha\beta,\mathbf k} c^{\dagger}_{\xi\alpha,\mathbf k}\,
	\mathcal{M}^{\phantom{\dagger}}_{\xi\alpha,\zeta\beta} \,
	c^{\phantom{\dagger}}_{\zeta\beta,\mathbf k}, \,\,
\end{eqnarray}
with the interaction matrix:
\begin{eqnarray}
	\label{eq:H-matrix}
	\mathcal{M}(\mathbf k) = \left(\begin{array}{ccccc}
	k_2\, k_3\,\mathbb{I}  &  & \bm\Delta_3\cdot\bm\sigma &  & \bm\Delta_2\cdot\bm\sigma \\
	\bm\Delta_3\cdot\bm\sigma &  & k_3\, k_1\,\mathbb{I}  & & \bm\Delta_1\cdot\bm\sigma \\
	\bm\Delta_2\cdot\bm\sigma &  & \bm\Delta_1\cdot\bm\sigma  & & k_1\, k_2\,\mathbb{I}
	\end{array}\right).
\end{eqnarray}
Here $\mathbb{I}$ is a $2\times2$ identity matrix, $\bm\sigma = (\sigma^x, \sigma^y, \sigma^z)$ 
is a vector of Pauli matrices, and we have set $m^*=1$ for simplicity.

To examine the structure of the SDW state near the ordering temperature, a Ginzburg-Landau expansion up to sixth order in $\bm\Delta$ was derived in Ref.~\cite{nandkishore12} by integrating out the fermions. The analysis found that
a collinear SDW state with $\bm\Delta_1 = \bm\Delta_2 = \bm\Delta_3$ is favored immediately below
the magnetic transition; the corresponding spin order is shown in Fig.~\ref{fig:sdw}(b).
At a lower temperature, the system undergoes another transition into a chiral SDW state whose three components of the
order parameter are orthogonal to each other and have the same amplitude, as shown in Fig.~\ref{fig:phase}(c).
The real-space magnetic order shown in Fig.~\ref{fig:sdw}(a) has a quadrupled unit cell with local spins pointing toward different
corners of a regular tetrahedron. An explicit calculation for the triangular-lattice Hubbard model gives a phase diagram shown in Fig.~\ref{fig:phase}(b), which is consistent with the two-stage ordering scenario described above.

\begin{figure}[t]
\includegraphics[width=0.94\columnwidth]{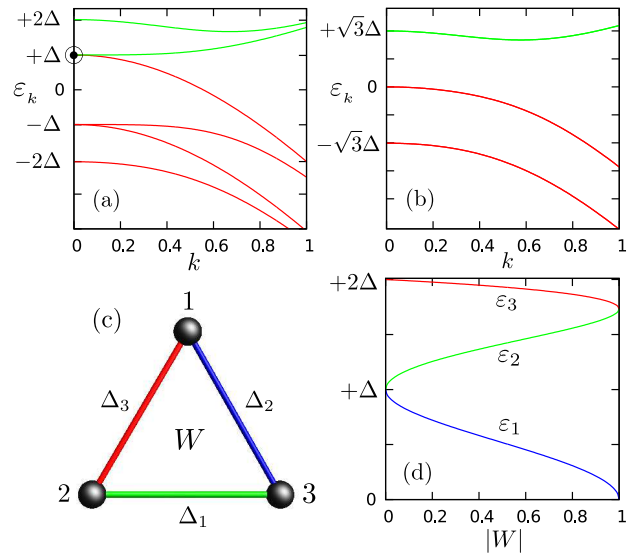}
\caption{(Color online) 
The mean-field band structure along $\mathbf k = (k, 0, 0)$ in the folded BZ for the (a) collinear 
and (b) noncoplanar SDW states. The calculation is done for the special case of triangular lattice.
The quadratic Fermi point with winding number $n = \pm 2$ is marked by $\odot$ in panel~(a).
Each band in the noncolanar SDW state is doubly degenerate. (c)~An equivalent tight-binding model on the 
triangular loop for Hamiltonian~(\ref{eq:H-matrix}). (d)~Energy levels at the $\mathbf k = 0$ as a function of the
invariant Wilson loop $W$. The collinear and tetrahedral SDW states have $|W| = 0$ and 1, respectively.
\label{fig:band}}
\end{figure}

The collinear SDW state is a half-metal which hosts a Fermi point at the center of the folded BZ; its mean-field
band structure is shown in Fig.~\ref{fig:band}(a).
To understand the nature of this gapless point, we first focus on the fermion spectrum at $\mathbf k=0$.
We assume that the three order parameters have the same amplitude $\Delta$, as such states
are favoured by the fourth order contributions to the energy expansion in powers of~$\bm\Delta_\eta$~\cite{nandkishore12}.
The interaction matrix~(\ref{eq:H-matrix}) at the $\Gamma$ point in {\em momentum} space is equivalent to a tight-binding
problem on a triangular loop (see Fig.~\ref{fig:band}(c)) with a spin-dependent hopping 
$t_{ij} = t\, { U}_{ij} = \bm\Delta_\eta\cdot\bm\sigma = \Delta\exp(i\frac{e}{h}\int \mathbf A\cdot d\mathbf r)$,
where $\mathbf A$ is a non-Abelian gauge potential. Its spectrum depends only on the 
gauge-invariant Wilson loop $W = \mbox{tr}\,(U_{12}\,U_{23}\,U_{31})$. 
Interestingly, this non-Abelian flux is proportional to the scalar spin chirality
\begin{eqnarray}
	\label{eq:W}
	W = -i \bm\Delta_1\cdot(\bm\Delta_2\times\bm\Delta_3)/\Delta^3.
\end{eqnarray}
Explicitly, the energy levels at the $\Gamma$ point are solutions of the
polynomial equation: $\varepsilon^2 (\varepsilon^2-3)^2 + 4(|W| - 1) = 0$, and always
appear in $\pm \varepsilon_m$ pairs. Fig.~\ref{fig:band}(d) shows the three positive eigen-energies
as a function of the non-Abelian flux.  Noting that there are three bands with negative 
energy, the charge gap at 2/3 filling is given by $\Delta\varepsilon = \varepsilon_2 - \varepsilon_1$.
This gap closes only when $W = 0$, i.e. for any coplanar, and particularly collinear, SDW state.

\begin{figure}[t]
\includegraphics[width=0.95\columnwidth]{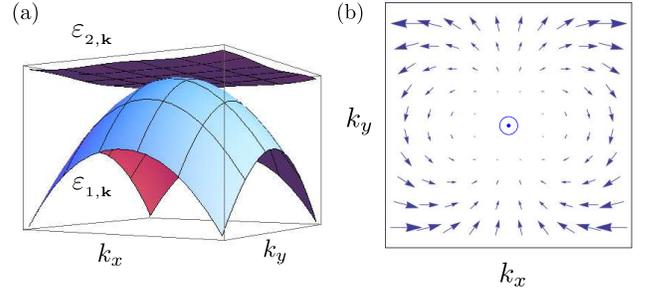}
\caption{(Color online) (a) Fermi point at a quadratic band crossing at $\mathbf k = 0$.  
(b)~The pseudovector field $\mathbf d = (d_x, d_z)$ with $d_x = (k_x^2 - k_y^2)/4$ and $d_z = k_x k_y/2$
resembles a vortex in XY systems with a winding number $n = 2$.
\label{fig:quadratic}}
\end{figure}

We now show that the transition from collinear to chiral SDW states is a topological phase transition
involving a quadratic Fermi point. Without loss of generality, we assume $\bm\Delta_1 = \bm\Delta_2 =
\bm\Delta_3 = \Delta\,{\hat{\mathbf z}}$ for the collinear order. 
The low-energy electrons in the vicinity of the emergent Fermi point
come from the two bands $\varepsilon_1$ and $\varepsilon_2$ in Fig.~\ref{fig:band}(d); the corresponding eigenstates are
$\psi_{1} = \frac{1}{\sqrt{2}}(c_{1 \downarrow} - c_{2 \downarrow})$
and $\psi_{2} = \frac{1}{\sqrt{6}}(c_{1 \downarrow} + c_{2 \downarrow} - 2c_{3 \downarrow})$, respectively. 
Introducing a pseudospin $\tau^z = \pm 1$ to label these
two states, the low-energy Hamiltonian in this basis is given by
\begin{eqnarray}
	\mathcal{H}(\mathbf k) = h_0\, \mathbb{I} + d_x \tau^x + d_z \tau^z,	
\end{eqnarray}
where $\tau^x$ and $\tau^z$ are Pauli matrices and
\begin{eqnarray}
	& & \qquad h_0 = \Delta - (k_x^2 + k_y^2)/4, \nonumber \\
	& & d_x = k_x k_y/2, \quad d_z = (k_x^2 - k_y^2)/4.
\end{eqnarray}
A diagonalization of  $\mathcal{H}(\mathbf k)$ yields a flat band $\varepsilon_{1,\mathbf k} = \Delta$
and a quadratic band $\varepsilon_{2,\mathbf k} = \Delta - |\mathbf k|^2/2$.
The pseudovector field $\mathbf d = (d_x, d_z)$ shown in Fig.~\ref{fig:quadratic}(b) has $d$-wave
symmetry. The topological charge for the singular $\mathbf k = 0$ point  is given by the winding
number of pseudovector field: $n = \frac{1}{2\pi} \oint_\mathcal{C} \nabla\theta_d(\mathbf k) \cdot d\mathbf k 
= \pm 2$ \cite{volovik03}, where $\theta_d = \arctan(d_z/d_x)$ and $\mathcal{C}$ is a contour enclosing the Fermi point.

The existence of this topological Fermi point is protected by the $C_6$ symmetry 
preserved in the collinear SDW phase and the spin collinearity masqueraded as an effective time-reversal symmetry.
Further symmetry-breaking at lower temperatures could remove the quadratic Fermi point
by either splitting it into elementary Dirac points or simply opening a gap. To study the stability
of the collinear SDW state, we introduce small deviations to the order parameters:
$\bm\Delta_\eta = \Delta\,{\hat{\mathbf z}} + \mathbf m_{\eta}$ with $\mathbf m_{\eta} \perp {\hat{\mathbf z}}$.
Substituting into Eq.~(\ref{eq:H_MF1}) and projecting to the low-energy doublet manifold $\Psi = (\psi_{1}, \psi_{2})$, the
mean-field Hamiltonian becomes
\begin{eqnarray}
	\label{eq:H_MF2}
	&&H_{\rm MF} = \frac{1}{v'}\int d\mathbf r \sum_{\eta} \left|\mathbf m_{\eta}\right|^2 
	+ \sum_{\mathbf k} \Psi^{\dagger}_{\mathbf k}\,\mathcal{H}(\mathbf k)\,\Psi^{\phantom{\dagger}}_{\mathbf k} \\
	&& \,\, + \int d\mathbf r\, \Psi^{\dagger}(\mathbf r)\!
	\left[\frac{1}{\sqrt{6}\Delta}\left(\mathcal{Q}_1\tau^z + \mathcal{Q}_2\tau^x\right) 
	+ \frac{\kappa}{\sqrt{3}\Delta^2}\, \tau^y \right]\!
	\Psi(\mathbf r), \nonumber
\end{eqnarray}
where $1/{v'} = 1/{v} + 1/6\Delta$ is the effective inverse coupling.
The doublet order parameter $(\mathcal{Q}_1, \mathcal{Q}_2)$ given by
\begin{eqnarray}
	\mathcal{Q}_1 &=& \bigl(\left|\mathbf m_1\right|^2 + \left|\mathbf m_2\right|^2 
	- 2 \left|\mathbf m_3\right|^2\bigr)/\sqrt{6}, \nonumber \\
	\mathcal{Q}_2 &=& \bigl(\left|\mathbf m_1\right|^2 - \left|\mathbf m_2\right|^2\bigr)/\sqrt{2},
\end{eqnarray}
describes a nematic phase for the $\Psi$ fermions in which the $C_6$ rotational symmetry is broken down to $C_2$ by splitting
the quadratic Fermi point into two Dirac points. The corresponding SDW order is dominated by a single nesting wavevector.
The nematic phase remains a half-metal.
Since the total winding number is conserved, the two residual Dirac points in the nematic 
phase carry the same topological charges. 
The order parameter $\kappa$ is the scalar spin chirality (\ref{eq:W}):
\begin{eqnarray}
	\kappa &=&  \bm\Delta_1\cdot(\bm\Delta_2\times\bm\Delta_3) = \Delta\bigl(m^x_1\,m^y_2 - m^y_1\, m^x_2 \\
	 & & +\, m^x_2\, m^y_3 - m^y_2\, m^x_3 + m^x_3\, m^y_1 - m^y_3\, m^x_1 \bigr). \nonumber
\end{eqnarray}
It characterizes an insulating phase with broken time-reversal symmetry and a zero-field quantized Hall 
conductivity $\sigma_{xy} = e^2/h$.

Minimization of Eq.~(\ref{eq:H_MF2}) yields a ground state that preserves the $C_6$ rotational symmetry
while breaking the effective time-reversal symmetry. The transition from the $\kappa = 0$ phase 
into the quantum Hall state with $\kappa \neq 0$ is a discontinuous one.
To see this, we compute the ground-state energy of~(\ref{eq:H_MF2}) as a function of $\kappa$ \cite{supp}:
\begin{eqnarray}
	\frac{E(\kappa)}{V} = \frac{2 \left|\kappa\right|}{\sqrt{3}\, v'\Delta} 
	- \frac{\rho^* \kappa^2}{6\Delta^4}\log\frac{\Lambda}{|\kappa|},
\end{eqnarray}
where $\rho^*$ is the density of states at the quadratic Fermi point, $\Lambda$ is an ultraviolet cutoff,
and $V$ is the system volume.
The function $E(\kappa)$ has two minima at $\kappa = 0$ and $\kappa \sim \Lambda/\sqrt{e}$.
At higher temperatures, the system may remain in the $\kappa = 0$ minimum corresponding to the collinear SDW state, as it was found in Ref.~\cite{nandkishore12} for 
a Hubbard model on triangular and honeycomb lattices.
Upon lowering the temperature, the system switches to the absolute minimum with $\kappa = \pm \Delta^3$ 
(since the triple product is bounded) which corresponds to the non-coplanar tetrahedral order.

The first-order transition scenario is consistent with our explicit mean-field calculations
of the Hubbard model on triangular or honeycomb lattices. It is worth noting that the collinear SDW
phase is stabilized by thermal fluctuations at finite temperatures. A lower free energy of the uniaxial SDW state results
from the gapless electronic excitations and the softer magnetic fluctuations associated with collinear spin order. 
Although the collinear SDW state hosting a topological Fermi point has been shown to exist in three of the most representative
2D lattices with $C_6$ rotational symmetry~\cite{nandkishore12,supp}, and maybe also in their decorated variants, 
its stability at finite temperatures depends on details of the model.

To summarize, we have shown the emergence of a quadratic Fermi point characterized by a winding number $n = \pm 2$
in  triple-$\mathbf Q$ collinear SDW states of different two-dimensional lattices. Such SDW phase arises from a perfectly nested
Fermi surface in lattices with a sixfold rotational symmetry such as triangular, honeycomb, or kagome.
Unlike most topological Fermi points which result from the special band structures of certain lattice problems,
this quadratic Fermi point emerges from a topologically trivial Fermi surface through electron interactions.
We have also obtained a universal low-energy Hamiltonian for the quadratic Fermi point and showed that
the instability towards a quantum Hall insulator is described by an order parameter which corresponds to the
scalar spin chirality. Our theory thus explains why the  topologically non-trivial (non-coplanar) SDW order  
is always more stable than the topologically trivial coplanar states at low temperatures.

Although spontaneous quantum Hall effect has been extensively discussed in the context of double-exchange or Kondo-lattice models,
most of these studies assume classical localized moments to begin with. It is then natural to ask if this phenomenon survives 
for models that incorporate quantum spin fluctuations. Our analysis of the noncoplanar
SDW ordering in Hubbard-like models provides a positive answer to this important question.
Possible material realizations of the collinear SDW phase that hosts the topological Fermi point 
include triangular compound Na$_{0.5}$CoO$_2$ \cite{ning08} and doped graphene~\cite{mcchesney10}.

{\it Acknowledgement.} We thank A.~Chubukov, R.~Fernandez, Y.~Kato, I.~Martin, and R.~Nandkishore
for useful discussions. Work at LANL was carried out under 
the auspices of the U.S.\ DOE contract No.~DE-AC52-06NA25396 
through the LDRD program. G.W.C. is grateful to the hospitality of CNLS at 
LANL and the support of ICAM and NSF grant DMR-0844115.

\newpage

\section{Supplementary materials}

\begin{figure*}[t]
\includegraphics[width=1.8\columnwidth]{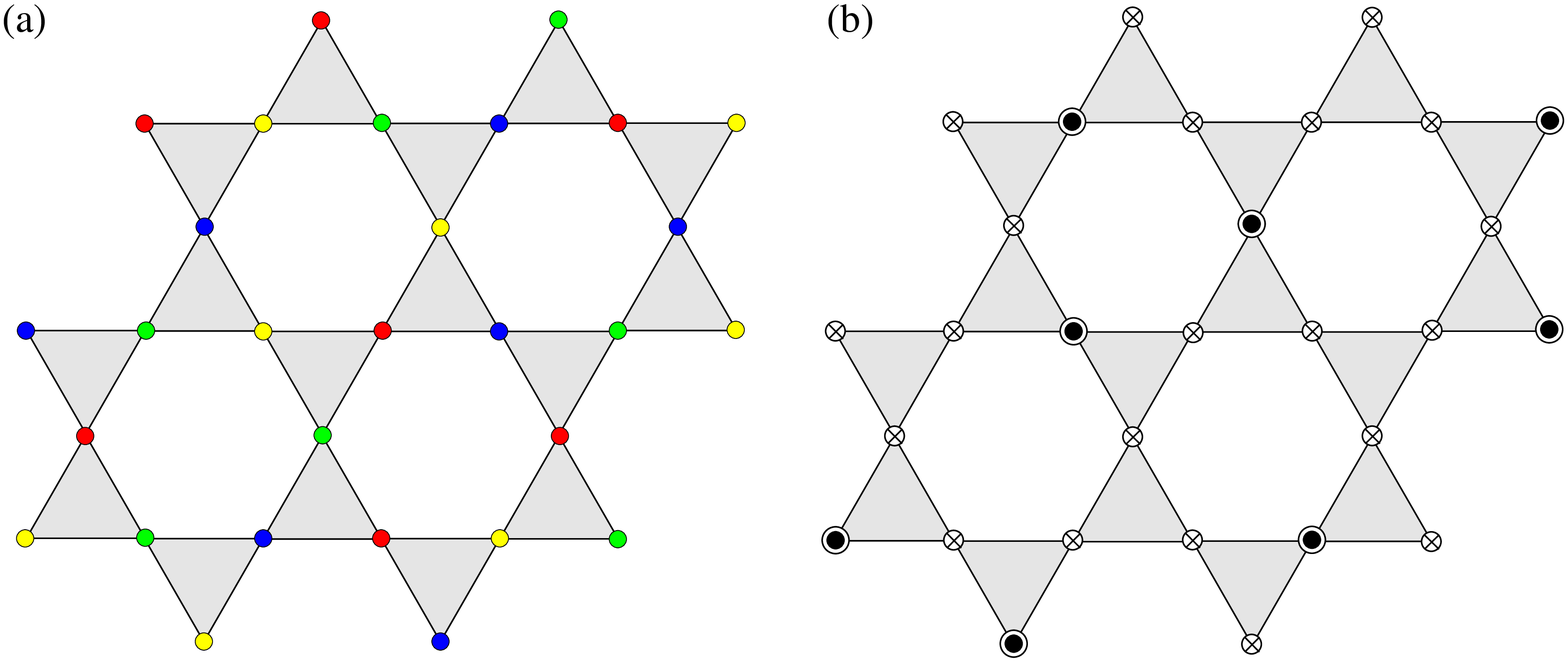}
\caption{(Color online) (a) The non-coplanar arrangement of local moments $\mathbf S(\mathbf r)$ on the
kagome lattice. Different colors correspond to the four non-coplanar spin directions pointing toward
the corners of a regular tetrahedron. (b) The collinear triple-$\mathbf Q$ SDW state on kagome. The local moments
at the $\odot$ and $\otimes$ sites are $3 \Delta$ and $- \Delta$, respectively. The quadrupled unit cell
in both SDW states contains 12 sites.
\label{fig:kagome}}
\end{figure*}

\section{Kagome lattice}

The triple-$\mathbf Q$ spin-density-wave (SDW) states in triangular and honeycomb lattices have been 
discussed in Refs.~\cite{martin08-s,li11-s,nandkishore12-s}. 
Here we consider another representative two-dimensional lattice with $C_6$ rotational symmetry: the case of kagome lattice. 
The existence and stability of non-coplanar tetrahedral SDW order on kagome lattice, shown in Fig.~\ref{fig:kagome}(a), 
has been demonstrated in Ref.~\cite{yu11-s}. The non-coplanar tetrahedral order shown in Fig.~\ref{fig:kagome}(a)
can be described as
\begin{widetext}
\begin{eqnarray}
	\mathbf S_\alpha(\mathbf r) &=& +\bm\Delta_1 \cos(\mathbf Q_1\cdot\mathbf r) + \bm\Delta_2 \cos(\mathbf Q_2\cdot\mathbf r) 
	+ \bm\Delta_3 \cos(\mathbf Q_3\cdot\mathbf r), \nonumber \\ 
	\mathbf S_\beta(\mathbf r) &=& +\bm\Delta_1 \cos(\mathbf Q_1\cdot\mathbf r) - \bm\Delta_2 \cos(\mathbf Q_2\cdot\mathbf r) 
	- \bm\Delta_3 \cos(\mathbf Q_3\cdot\mathbf r), \\ 
	\mathbf S_\gamma(\mathbf r) &=& -\bm\Delta_1 \cos(\mathbf Q_1\cdot\mathbf r) - \bm\Delta_2 \cos(\mathbf Q_2\cdot\mathbf r) 
	+ \bm\Delta_3 \cos(\mathbf Q_3\cdot\mathbf r), \nonumber   
\end{eqnarray}
\end{widetext}
where $\alpha$, $\beta$, and $\gamma$ denote the three sublattices, and vector order parameters $\bm\Delta_{1,2,3}$ have
the same amplitude and are orthogonal to each other. 
Since the unit cell of kagome lattice contains three inequivalent sites, 
there are several triple-$\mathbf Q$ collinear SDW orderings with different arrangements of spins within the unit cell.
The collinear SDW state that is topologically connected to the tetrahedral order is obtained from Eq.~(1) above by 
continuously closing the solid angle between the three vector order parameters $\bm\Delta_i$.
By setting $\bm\Delta_1 = \bm\Delta_2 = \bm\Delta_3 = \Delta \hat{\mathbf z}$, the resulting collinear SDW state is shown 
in Fig.~\ref{fig:kagome}(b) in which 3 sites in the quadrapuled unit cell have magnetic moment $3 \Delta$ while
the moment at the remaining 9 sites is $- \Delta$.

\begin{figure*}
\includegraphics[width=1.98\columnwidth]{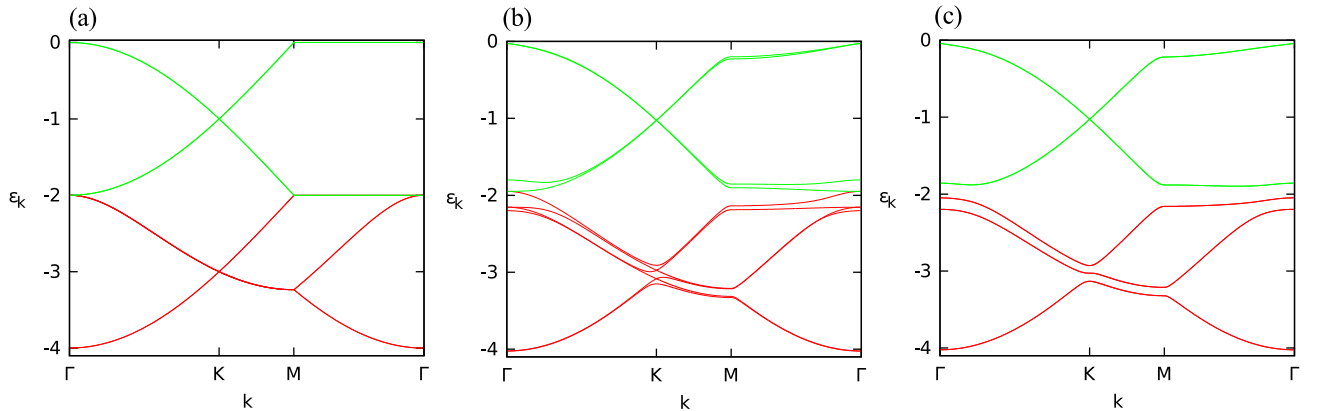}
\caption{(Color online) (a) Tight-binding band structure of kagome lattice shown with the folded Brillouin zone. 
(b) and (c) are the mean-field band structures of kagome Hubbard model in the presence of collinear and non-coplanar (tetrahedral) SDW ordering,
respectively. At filling fraction $\nu = 1/4$, there exists a quadratic Fermi point at the $\Gamma$ point in the collinear SDW state, while
the spectrum is gapped in the non-coplanar SDW state.
\label{fig:ek}}
\end{figure*}

We now consider the Kondo-lattice model, which can be viewed as the mean-field approximation of the Hubbard-like model,
on the kagome lattice:

\begin{eqnarray}
	H = -t \sum_{\langle ij \rangle} c^{\dagger}_{i,\alpha} c^{\phantom{\dagger}}_{j,\alpha} - J_H \sum_i \mathbf S_i \cdot 
	c^{\dagger}_{i,\alpha}\bm\sigma_{\alpha\beta} c_{j,\beta}.
\end{eqnarray}
where $\mathbf S_i = \frac{1}{2}\langle c^{\dagger}_{i,\alpha}\bm\sigma_{\alpha\beta} c_{j,\beta} \rangle$ is the local magnetic moment.
The band structure of the nearest-neighbor tight-binding model ($J_H = 0$) is shown in Fig.~\ref{fig:ek}(a). A logarithmically 
divergent density of states occurs at 1/4-filling due to perfectly nested Fermi surface. Figs.~\ref{fig:ek}(b) and (c)
shows the mean-field band structures in the presence of collinear and tetrahedral SDW order, respectively.
At 1/4-filling, the lowest 6 bands are completely filled, giving rise to a quadratic Fermi point at $\mathbf k=0$ ($\Gamma$ point
in the folded Brillouin zone) [Fig.~\ref{fig:ek}(b)]. On the other hand, the non-coplanar tetrahedral order
opens a charge gap as shown in Fig.~\ref{fig:ek}(c).

With the addition of the kagome case discussed here, 
we have demonstrated the existence of triple-$\mathbf Q$ collinear SDW state and its topological
connection to the non-coplanar SDW order in three of the most representative 2D lattices with $C_6$ symmetry, namely triangular, 
honeycomb, and kagome lattices.
Although perfect Fermi surface nesting and divergent Van Hove singularities also exist in other lattices such as the decorated
variants of the above three lattices, the properties of the triple-$\mathbf Q$ SDW ordering will be left for future studies.

\section{Ground-state energy}

\begin{figure}
\includegraphics[width=0.95\columnwidth]{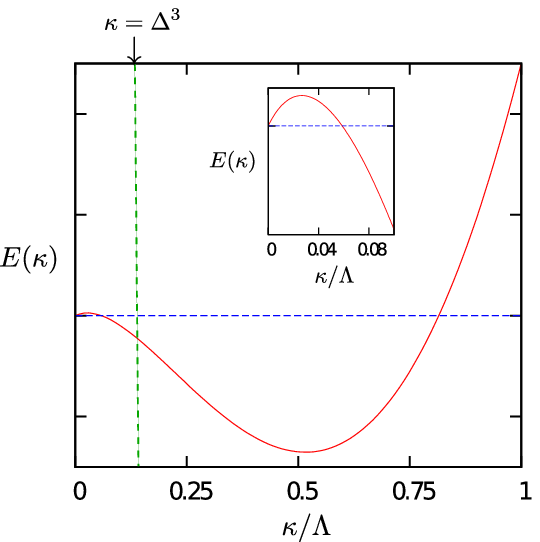}
\caption{(Color online) Ground-state energy $E(\kappa)$ as a function of the spin scalar 
chirality $\kappa =  \bm\Delta_1\cdot(\bm\Delta_2\times\bm\Delta_3)$.
\label{fig:E_gs}}
\end{figure}

In this section we discuss the derivation of ground state energy Eq.~(10) in the main text.
\begin{eqnarray}
	\frac{E(\kappa)}{V} = \frac{2 \left|\kappa\right|}{\sqrt{3}\, v'\Delta} 
	- \frac{\rho^* \kappa^2}{6\Delta^4}\log\frac{\Lambda}{|\kappa|},
\end{eqnarray}
The first term is the energy cost when spins deviate from the collinear limit. 
It comes from the term $\frac{1}{v'}\sum_{\eta}|\mathbf m_\eta|^2$ in the effective Hamiltonian Eq.~(7) of the main text. 
As also discussed there, the ground state preserves the $C_6$ rotational symmetry, hence the
nematic order parameter vanishes $\mathcal{Q}_1 = \mathcal{Q}_2 = 0$. The scalar spin chirality is then maximized when
the three $\mathbf m_\eta$ vectors lie in the same plane perpendicular to $\hat{\mathbf z}$ and point at 120$^\circ$ to one another.
This gives rise to $\sum_\eta |\mathbf m_\eta|^2 = 2|\kappa|/\sqrt{3}\Delta$.

The presence of spin chirality $\kappa$ opens a gap at the Fermi point. The dispersion of the two
electron bands near the $\Gamma$ point is: $\epsilon_{1,2}(\mathbf k) \approx h_0(\mathbf k) 
\pm \sqrt{|\mathbf k|^4 + \delta^2}$, where the energy gap $\delta = |\kappa|/\sqrt{3}\Delta^2$.
Standard calculation gives an energy gain $\int d^2\mathbf k  (\sqrt{|\mathbf k|^4 + \delta^2} - |\mathbf k|^2 )
\sim \kappa^2 \log(\Lambda/|\kappa|)$ which is logarithmically divergent with $\kappa$. Here $\Lambda$ denotes an ultraviolet cutoff.
This accounts for the second term in Eq.~(3).

Fig.~\ref{fig:E_gs} shows the ground state energy $E(\kappa)$ as a function of the scalar chirality $\kappa$ 
for a generic case. The first term in Eq.~(3) which is linear in $|\kappa|$ gives rise to a local minimum at $\kappa = 0$,
corresponding to the collinear SDW state. The function $E(\kappa)$ has a global minimum at $\kappa \sim \Lambda/\sqrt{e}$
coming from the second term in Eq.~(3). However, since the scalar chirality is bounded $|\kappa| \le \Delta^3$, the ground
state is reached when the three order parameters $\bm\Delta_\eta$ are perpendicular to each other, corresponding to 
the non-coplanar tetrahedral SDW order.


\begin{thebibliography}{99}
\bibitem{note1} Quadratic Fermi point with zero winding number is an accidental band crossing and can be removed by small
perturbations without breaking any symmetries.
\end{thebibliography}
\end{document}